\documentstyle{article}

\def\be{\begin{equation}}
\def\ee{\end{equation}}
\def\bea{\begin{eqnarray}}
\def\eea{\end{eqnarray}}
\newcommand{\tr}[1]{\,{\rm tr}\,#1\,}

\begin{document}

\title{D-branes, Black Holes and  $SU(\infty)$ Gauge Theory}

\author{
 I.V.Volovich \\
Steklov Mathematical Institute,\\
Vavilov st. 42, GSP-1, 117966, Moscow, Russia}
\date {$~$}
\maketitle

\begin {abstract}

We discuss an application of the  known in QCD large $N$ 
expansion to strings and supermembranes in the strong  coupling. 
In particular we use the recently obtained master field describing 
$ SU(\infty)$ gauge theory to argue that quantum extreme 
black holes obey quantum Boltzmann (infinite) statistics. This 
supports a topological argument by Strominger that black holes 
obey infinite statistics. We also speculate on a formulation of 
$X$-theory of strings and p-branes as theory of Grothendieck's 
motives.  The partition function is expressed in terms
of $L$-function of a motive. The Beilinson conjectures on the
values of $L$-functions are interpreted as dealing with the 
cosmological constant problem.
\end {abstract}

Duality connects week and strong coupling regimes in string theory 
\cite {FIL}.
We know how to deal with the week coupling regime but one needs a 
nonperturbative method to control the strong coupling
regime. Actually one knows  the only one systematic nonperturbative
method  in quantum field theory. This is the large $N$ expansion.
In this talk we shall discuss some applications of the large $N$ expansion
to theory of superstrings and $p$-branes.

Recently a remarkable progress towards understanding of the
black hole entropy in superstring theory has taken place.
Strominger and Vafa  \cite {SV} used the Dirichlet (D)-brane approach
\cite {Pol} to identify  and to count the degenerate
quantum states which have the same quantum numbers
of certain extreme black holes.  They have shown that the growth of the
elliptic genus of an appropriate Kahler manifold agrees 
 with the Bekenstein-Hawking entropy 
formula  in the limit of large charges.

Quantum states describing black hole in the D-brane approach
consist of a large number $N$ of D-branes. Witten \cite {Wit1}
has shown that the effective action of
$N$ parallel coincident Dirichlet $p$-branes is the dimensional
reduction  of the ten-dimensional $U(N)$ supersymmetric Yang-Mills theory
to $p+1$-dimensions. Therefore it seems  natural 
to use the known in QCD  large $N$ expansion \cite {GG,Dou,AV} 
to study quantum black holes.
The large $N$ expansion works even in the strong coupling
regime  therefore in principle one can apply it to large
black holes when there are problems how to deal with strong
interactions \cite {MS}.
For $0$-brane case one gets a model of the $SU(N)$ supersymmetric
gauge quantum mechanics.  In the limit $N\rightarrow
\infty$ this model yields the supermembrane action and the gauge
 group $ SU(\infty)$
approaches the group of area-preserving transformations corresponding to 
a membrane with topology of sphere \cite{DHH,Tow}.

Here we shall argue that the Hilbert space for extreme black holes
and membranes is the Boltzmannian Fock
space defined by the following relations for creation and annihilation
operators:
\begin{equation}
	a_{k}a_{l}^{+}=\delta_{kl}
\end{equation}
(this is $q$-deformed commutational relation  
$a_{k}a_{l}^{+}-qa_{l}^{+}a_{k}=\delta_{kl}$ for $q=0$).
This follows from the general result
obtained in   \cite{AV} that the colourless sector of
 $ SU(\infty)$ invariant
theory in $d$-dimensional ($d\geq 1$) spacetime is described by states
in the Fock space defined by relations (1). For 
 $ SU(N)$ gauge field $A_{\mu}(x)$
in $d$-dimensional
Minkowski space-time  one has the
following basic relation
$$\lim _{N \to \infty}
\frac{1}{N^{1+\frac{n}{2}}}<0|\tr(A_{\mu_{1}}(x_1)...A_{\mu_{n}}(x_n))|0>
=(\Omega_{0}|B_{\mu_{1}} (x_{1})...B_{\mu_{n}} (x_{n})|\Omega_{0})
$$
Here the limiting field $B_{\mu}(x)$,  so called master field,
is defined  as a solution of the Yang-Feldman equation
\begin{equation}
B_{\mu}(x)=B_{\mu}^{(in)}(x)+\int D^{ret}_{\mu\nu}(x-y)J_{\nu} (y)dy ,
\nonumber
\end{equation}
where the free field $B_{\mu}^{(in)}(x)$ is quantized according
to the rule (1).
Here $|0>$ is vacuum in the ordinary Bosonic Fock space and $|\Omega_{0})$
is vacuum in the Boltzmannian Fock space.
The master field  $B_{\mu}(x)$ does not have matrix indexes. It satisfies
to the ordinary Yang-Mills equations  but
it is quantized according to the new rule. 
 The gauge group for the gauge master field
is an infinite dimensional group of unitary operators in the Boltzmannian 
Fock space. After the fixing the gauge one gets the BRST symmetry.
The gauge  group is approached by the group of area-preserving 
diffeomorfisms of the membrane.

BPS-saturated states in spacetime correspond to BPS-states of the D-brane 
worldvolume theory. One interprets the BPS-states as a black hole in 
space-time only for large charges, i.e. in the large $N$ limit. But this 
limit is described by quantum theory in the Boltzmannian Fock space, as
it has been discussed earlier. Thus quantum  states of extreme black hole 
belong to the Boltzmannian Fock space. This supports a topological 
argument made by Strominger \cite{Str} that extreme black holes obey infinite
statistics. He argued that black hole exchange is the exchange of
two wormhole ends. This is not a diffeomorphism and therefore
extremal black hole scattering resembles that of distinguishable particles.

Duality and  D-branes led to a  new insight into the structure
of superstring theory but also pointed up that our current understanding
of string theory is only valid up to some scale. Probably string theory
can not be fundamentally defined as a theory of extended objects of any
kind - including strings \cite{BBS}. It was suggested  \cite{Vol1,Vol2} that
theory of 
Grothendieck's motives \cite{Man1,Ser,RSS} can be useful in attempts
 to find a fundamental
"$X$-theory" of strings and p-branes. Motives are defined by algebraic 
correspondences modulo homological equivalence. Motivic cohomology
is a kind of universal  cohomology theory for algebraic varieties.
Realizations of a motive  $M$ over the field of rational numbers $Q$
  are linear spaces over $Q$
 and over the field of $l$-adic numbers $Q_{l}$. If $X$ is
 a smooth projective algebraic variety over $Q$ then the realization
 of its motive is given by the Betti $H_{B}(X)$, De Rham $H_{DR}(X)$
 and $l$-adic cohomology $H_{l}(X)$ of $X$. Category of motives is 
 isomorphic to category of finite dimensional representations
 of proalgebraic motivic Galois-Serre group $G$. In the 
  "$X$-theory" the group $G$ plays the role of
 conformal group in the ordinary string theory.
 Motivic $X$-theory deals with algebraic varieties over the field of 
 rational numbers $Q$, so the theory is background independent
 and it is not based on a spacetime continuum.
 The partition function in $X$-theory is given by $L$-function of a motive.
 $L$-function
 of a motive $M$ is defined by the Euler product:
 $ L(M,s)=\prod_{p}det(1-p^{-s}Frob_{p}|H_{l}(M)^{I_{p}})^{-1}$,
where $Frob_{p}$ is a Frobenius element in $G_{Q_{p}}$ and $I_{p}$
is the inertia group in  $G_{Q_{p}}$. One assumes the existence
of the meromorphic continuation and  functional equation for $	L(M,s)$.

String  partition function  can be expressed  as inverse to the Mellin
transform of  $L$-function of a motive.
  Motive $M$  of bosonic strings has been
constructed by  Deligne and it was used in the proof of the Ramanujan  
conjecture: $|\tau(p)|\leq 2p^{11/2}$ \cite{Ser,Del}.
 The Ramanujan function $\tau (n)$ is defined by the
relation
$\Delta (q)=
q\prod_{n}(1-q^{n})^{24}=\sum_{n}\tau (n)q^{n}$
One has: $ \Delta (q)^{-1}=tr q^{H}$ where $H=L_{0}-1/24$ and
 $L_{0}$ is the Virasoro operator. $L$-function of the motive
 $M$ is the Dirichlet series:
\begin{equation}
L(M,s)=\sum_{n}\tau 
(n)n^{-s}=\prod_{p}(1-\tau(p)p^{-s}-p^{11-2s})^{-1},
\nonumber
\end{equation}

It is amusing that the motive $M$ is eleven-dimensional.
Theory of motives shows geometrical aspects
of modular functions widely used in string theory \cite{HM,DVV,Kaw}

To understand the vanishing of the cosmological constant we have
to consider values of  partition function. In $X$-theory
this means the investigation of values of $L$-functions of  motives. Therefore
the mystery of the cosmological constant is ultimately reduced
 to the Beilinson conjectures \cite{RSS}
on values of $L$-functions.

For a further discussion of arithmetical  physics see \cite{VVZ,Man2,FW,Jul,BB,Vol3}
$$~$$

This work was partially supported by  Grant RFFI-93-011-140.

\end{document}